\documentclass[twocolumn]{aastex62}
\usepackage{amsmath}
\usepackage{float}
\graphicspath{{./}{figures/}}

\shorttitle{Interstellar Meteors}

\begin{document}

\title{Probing Extrasolar Planetary Systems with Interstellar Meteors}

\email{amir.siraj@cfa.harvard.edu, aloeb@cfa.harvard.edu}

\author{Amir Siraj}
\affil{Department of Astronomy, Harvard University, 60 Garden Street, Cambridge, MA 02138, USA}

\author{Abraham Loeb}
\affiliation{Department of Astronomy, Harvard University, 60 Garden Street, Cambridge, MA 02138, USA}




\begin{abstract}
The first interstellar object, `Oumuamua, was discovered in the Solar System by Pan-STARRS in 2017, allowing for a calibration of the impact rate of interstellar meteors of its size $\sim 100\;$m. The discovery of CNEOS 2014-01-08 allowed for a calibration of the impact rate of interstellar meteors of its size $\sim 1\;$m. Analysis of interstellar dust grains have allowed for calibrations of the impact rate of smaller interstellar meteors down to the size $\sim 10^{-8}\;$m. We analyze the size distribution of interstellar meteors, finding that for smooth power-law fits of the form $N(r)\propto r^{-q}$, the possible values of $q$ are in the range $3.41 \pm 0.17$. We then consider the possibility of analyzing interstellar meteors to learn about their parent planetary systems. We propose a strategy for determining the orbits and chemical compositions of interstellar meteors, using a network of $\sim 600$ all-sky camera systems to track and conduct remote spectroscopy on meteors larger than $\sim 5$cm once every few years. It should also be possible to retrieve meteorites from the impact sites, providing the first samples of materials from other planetary systems.
\end{abstract}



\section{Introduction}
`Oumuamua was the first interstellar object detected in the Solar System by Pan-STAARS \citep{Meech2017, Micheli2018}. Several follow-up studies of `Oumuamua were conducted to better understand its origin and composition \citep{Bannister2017, Gaidos2017, Jewitt2017, Mamajek2017, Ye2017, Bolin2017, Fitzsimmons2018, Trilling2018, Bialy2018, Hoang2018, Siraj2019a, Siraj2019b, Seligman2019}. `Oumuamua's size was estimated to be 20m - 200m, based on Spitzer Space Telescope constraints on its infrared emission given its expected surface temperature based on its orbit \citep{Trilling2018}.

There is significant evidence for previous detections of interstellar meteors \citep{Baggaley1993, Hajdukova1994, Taylor1996, Baggaley2000, Mathews1998, Meisel2002a, Meisel2002b, Weryk2004, Afanasiev2007, Musci2012, Engelhardt2017, Hajdukova2018, Siraj2019c}. CNEOS 2014-01-08 is the largest interstellar meteor discovered in the Solar System \citep{Siraj2019c}. Spectroscopy of gaseous debris of interstellar meteors as they burn up in the Earth's atmosphere could reveal their composition \citep{Siraj2019c}. In this \textit{Letter}, we explore the size distribution of interstellar meteors and motivate the investigation of interstellar meteors as a new branch of astronomical research. We present a strategy for conducting spectroscopy and obtaining physical samples of interstellar meteors.

\section{Size Distribution}
\label{sec:methods}

\begin{figure*}
  \centering
  \includegraphics[width=\linewidth]{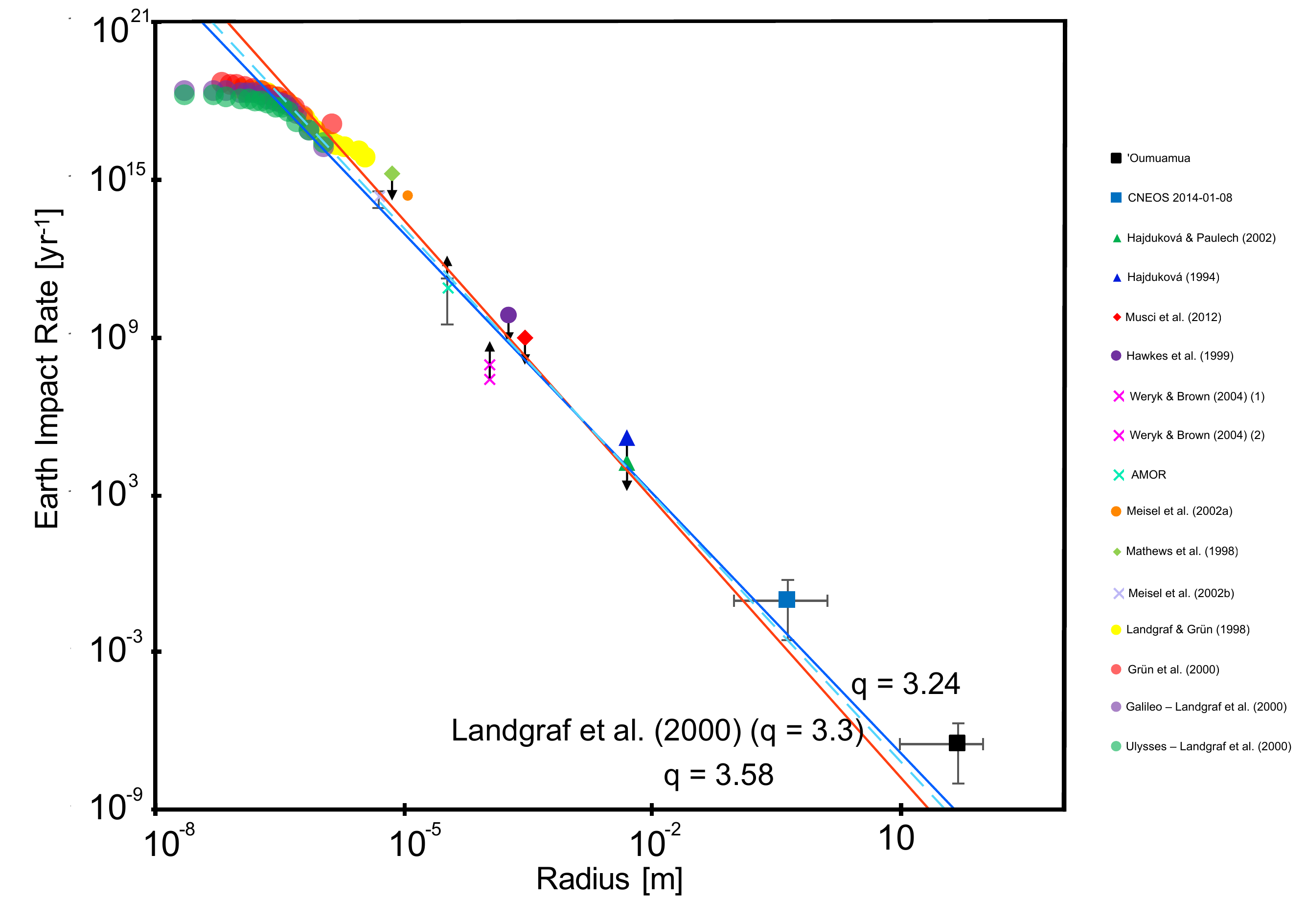}
    \caption{Interstellar meteoroid impact rate estimates from various studies, compiled by \cite{Musci2012}, with CNEOS 2014-01-08 and `Oumuamua added. The range of possible values for the \cite{Meisel2002b} measurement are for Geminga supernova particles assuming different models and fits. The AMOR data include results from \cite{Baggaley1993}, as well as the interpretation of those data from \cite{Taylor1996} and \cite{Baggaley2000}. The \cite{Weryk2004} values are for 2 and 3 standard deviations, respectively, above the hyperbolic speed limit. The power law fits that are consistent with all measurements for $r > 5 \times 10^{-6} \; \mathrm{m}$ (except for the controversial \cite{Meisel2002a} result) range from q = 3.24 to q = 3.58, consistent with the value of q = 3.3 inferred by \cite{Landgraf2000}.}
    \label{fig:1}
\end{figure*}

CNEOS 2014-01-08 and `Oumuamua serve as important calibration points for the size distribution of interstellar meteors, included with the results compiled by \cite{Musci2012} in Fig.~\ref{fig:1}. Non-detections from \cite{Hajdukova1994}, \cite{Hawkes1999}, \cite{Hajdukova2002}, and \cite{Musci2012} serve as upper limits. Detections from \cite{Weryk2004}, \cite{Baggaley1993} serve as lower limits, and the range of values given by \cite{Meisel2002b} are from different models and fits for Geminga supernova particles. \cite{Mathews1998} reported 1 and \cite{Meisel2002a} reported 143 detections from Arecibo, but the results are controversial due to large velocity uncertainties \citep{Musci2012}.

The possible smooth power-law fits of the form $N(r)\propto r^{-q}$ for all $r > 5 \times 10^{-6} \; \mathrm{m}$ except for the controversial \cite{Meisel2002a} result range from $q = 3.24$ to $q = 3.58$, consistent with $q = 3.3$ from \cite{Landgraf2000} but not with the power-law fits from \cite{Mathis1977}, \cite{Landgraf1998}, or \cite{Hajdukova2002}.

\section{Detection Strategy}
\label{sec:results}

The cores of meteoroids with radii larger than $\sim 5$cm can reach the ground in the form of meteorites \citep{Kruger2014}. Additionally, meteoroids on smaller size scales could be accelerated from the Poynting-Robertson effect and could have potential origins in the interstellar medium. Hence, interstellar meteors above this size are optimal for a systematic study of physical extrasolar material (in addition to the spectroscopy of the hot gases as the meteor burns up). Since we expect interstellar meteors of this size to strike the Earth a few times per year, a network of all-sky camera systems monitoring the sky above all land on Earth could detect an interstellar meteor of this size every few years. Such detections can be made with science-grade video cameras, such as those used in AMOS, CAMO, and CAMS\footnote{http://cams.seti.org/} \citep{Toth2015, Weryk2013}.

A conservative estimate for the total area of $\sim70$km altitude atmosphere visible from a system of two all-sky camera systems separated by 100km is $5 \times 10^5 \; \mathrm{km^2}$, so to cover all land on Earth would require $\sim 300$ systems, or $\sim 600$ total all-sky camera systems, similar to CAMO but with an all-sky field of view, like AMOS and CAMS\footnote{http://cams.seti.org/} \citep{Weryk2013, Toth2015}.

We therefore advocate for a network of all-sky camera systems to conduct real-time remote spectroscopy of the hot gases as $r \geq 5 \times 10^{-2}\; \mathrm{m}$ interstellar meteors burn up, and to precisely determine their trajectories for the immediate retrieval of interstellar meteorite samples.

\section{Discussion}
\label{sec:discussion}

We analyzed the updated size distribution of interstellar meteors, deriving a range of possible slopes of $q = 3.41 \pm 0.17$, consistent with the slope calculated by \cite{Landgraf2000}. We then presented a strategy for studying interstellar meteors: using a network of $\sim 600$ all-sky camera systems to determine the orbits and chemical compositions of $r \geq 5$cm meteors. This method also allows for the possibility of retrieving interstellar meteorite samples.

By extrapolating the trajectory of each meteor backward in time and analyzing the relative abundances of each meteor's chemical isotopes, one can match meteors to their parentt stars and reveal insights into planetary system formation. R-processed elements, such as Eu, can be detected in the atmospheres of stars \citep{Frebel2016}, so their abundances in meteor spectra can serve as important links to parent stars. This new field of astronomical research is significant as it would save the trip and allow us to study samples of materials from other planetary systems, be it natural or artificial in origin.

\section*{Acknowledgements}
We thank Peter Veres for helpful comments on the manuscript. This work was supported in part by a grant from the Breakthrough Prize Foundation. 






\begin{figure*}
  \centering
  \includegraphics[width=\linewidth]{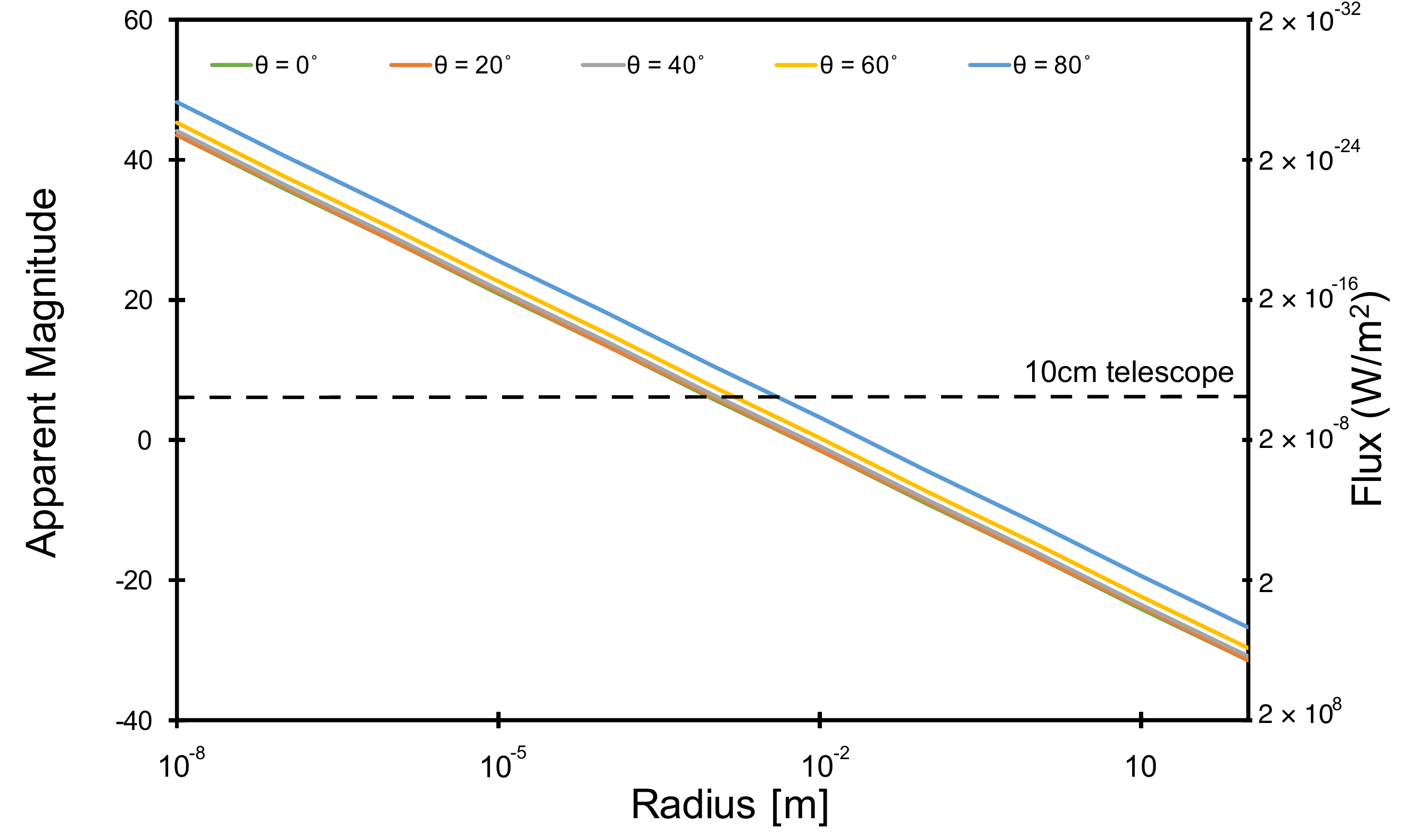}
    \caption{Apparent visual magnitude for $5 \times 10^{-2}$ s exposures as a function of meteor size at an altitude of 100 km, with $\theta$ measured from zenith to horizon. We assume a fiducial density of $\rho = 2\times 10^3 \;\mathrm{kg \; m^{-3}}$, an impact speed of $v_{impact} = 30 \; \mathrm{km \; s^{-1}}$, and that the meteor lifetime is on the order of a few seconds. Dashed lines the flux limit for a 10-$\sigma$ detection in each spectral bin with a spectral resolution of $R=10^3$, for telescope apertures of diameter 10cm and 1m.}
    \label{fig:2}
\end{figure*}



\end{document}